\input harvmac

\input epsf.tex
\ifx\epsfbox\UnDeFiNeD\message{(NO epsf.tex, FIGURES WILL BE IGNORED)}
\def\figin#1{\vskip2in}
\else\message{(FIGURES WILL BE INCLUDED)}\def\figin#1{#1}\fi
\def\ifig#1#2#3{\xdef#1{fig.~\the\figno}
\midinsert{\centerline{\figin{#3}}%
\smallskip\centerline{\vbox{\baselineskip12pt
\advance\hsize by -1truein\noindent{\bf Fig.~\the\figno:} #2}}
\bigskip}\endinsert\global\advance\figno by1}

\noblackbox
%
%
\font\tenbifull=cmmib10 
\font\tenbimed=cmmib10 scaled 800
\font\tenbismall=cmmib10 scaled 666
\textfont9=\tenbifull \scriptfont9=\tenbimed
\scriptscriptfont9=\tenbismall

\def\IZ{\relax\ifmmode\mathchoice
{\hbox{\cmss Z\kern-.4em Z}}{\hbox{\cmss Z\kern-.4em Z}}
{\lower.9pt\hbox{\cmsss Z\kern-.4em Z}}
{\lower1.2pt\hbox{\cmsss Z\kern-.4em Z}}\else{\cmss Z\kern-.4em Z}\fi}

\font\cmss=cmss10 \font\cmsss=cmss10 at 7pt

\font\ttsmall=cmtt10 at 8pt

\def\yboxit#1#2{\vbox{\hrule height #1 \hbox{\vrule width #1
\vbox{#2}\vrule width #1 }\hrule height #1 }}
\def\fillbox#1{\hbox to #1{\vbox to #1{\vfil}\hfil}}
\def\ybox{{\lower 1.3pt \yboxit{0.4pt}{\fillbox{8pt}}\hskip-0.2pt}}

\def\comments#1{}

\def\CN{{\cal N}}

\def\a{\alpha}

\def\II{\relax{I\kern-.07em I}}

\def\inbar{\,\vrule height1.5ex width.4pt depth0pt}
\def\IZ{\relax\ifmmode\mathchoice
{\hbox{\cmss Z\kern-.4em Z}}{\hbox{\cmss Z\kern-.4em Z}}
{\lower.9pt\hbox{\cmsss Z\kern-.4em Z}}
{\lower1.2pt\hbox{\cmsss Z\kern-.4em Z}}\else{\cmss Z\kern-.4em
Z}\fi}
\def\IB{\relax{\rm I\kern-.18em B}}
\def\IC{{\relax\hbox{$\inbar\kern-.3em{\rm C}$}}}
\def\ID{\relax{\rm I\kern-.18em D}}
\def\IE{\relax{\rm I\kern-.18em E}}
\def\IF{\relax{\rm I\kern-.18em F}}
\def\IG{\relax\hbox{$\inbar\kern-.3em{\rm G}$}}
\def\IGa{\relax\hbox{${\rm I}\kern-.18em\Gamma$}}
\def\IH{\relax{\rm I\kern-.18em H}}
\def\IK{\relax{\rm I\kern-.18em K}}
\def\IP{\relax{\rm I\kern-.18em P}}

\font\cmss=cmss10 \font\cmsss=cmss10 at 7pt
\def\IR{\relax{\rm I\kern-.18em R}}

%


%

\def\NP{{\it Nucl. Phys.\ }}

\def\PL{{\it Phys. Lett.\ }}
\def\PR{{\it Phys. Rev.\ }}
\def\PRL{{\it Phys. Rev. Lett.\ }}

%
\lref\joe{ J. Polchinski, \PRL {\bf 75} (1995) 4724, hep-th/9510017 and
``TASI Lectures on D-branes,'' preprint NSF-ITP-96-145 hep-th/9611050.}
\lref\gipo{E. Gimon and J. Polchinski, ``Consistency Conditions for
Orientifolds and D-Manifolds,'' \PR {\bf D54} (1996) 1667, hep-th/9601038.}
\lref\egk{S. Elitzur, A. Giveon and D. Kutasov, ``Branes and N=1
Duality In String Theory,'' hep-th/9702014.}
\lref\egkrs{S. Elitzur, A. Giveon, D. Kutasov, E. Rabinovici and
A. Schwimmer, ``Brane Dynamics and N=1
Supersymmetric Gauge Theory,''  hep-th/9704104.}
\lref\ejs{ N. Evans, C. V. Johnson and A. Shapere, ``Orientifolds,
Branes and Duality of 4D Gauge Theories,'' hep-th/9703210.}
\lref\bsty{A. Brandhuber, J. Sonnenschein, S. Theisen and S.
Yankielowicz, ``Brane Configurations and 4D Field Theories,'' hep-th/9704044.}
\lref\supered{E. Witten, ``Solutions of four-dimensional Field
Theories via M-Theory,'' hep-th/9703166.}
\lref\swi{N. Seiberg and E. Witten, ``Monopole Condensation, And Confinement
In N=2 Supersymmetric Yang-Mills Theory,'' \NP {\bf 426} (1994) 19,
hep-th/9407087.}
\lref\swii{N. Seiberg and E. Witten, ``Monopoles, Duality and Chiral
Symmetry Breaking in N=2
Supersymmetric QCD,'' \NP {\bf 431} (1994) 484, hep-th/9408099.}
\lref\klty{A. Klemm, W. Lerche, S. Theisen, and S. Yankielowicz, ``Simple
Singularities and N=2 Supersymmetric Yang-Mills,''
\PL {\bf B344} (1995) 169, hep-th/9411048.}
\lref\arfa{P. Argyres and A. Faraggi, ``The Vacuum Structure and Spectrum of
N=2 Supersymmetric SU(N) Gauge Theory,'' \PRL {\bf 73} (1995) 3931,
hep-th/9411057.}
\lref\haoz{A. Hanany and Y. Oz, ``On the Quantum Moduli Space of Vacua
of N=2 Supersymmetric $SU(N_c)$ Gauge Theories,''
\NP {\bf B452} (1995) 283, hep-th/9505075.}
\lref\aps{P. Argyres, M. Plesser, and A. Shapere, ``The Coulomb Phase of
N=2 Supersymmetric QCD,'' \PRL {\bf 75}(1995)
1699, hep-th/9505100.}
\lref\dasu{U. Danielsson and B. Sundborg, ``The Moduli Space and
Monodromies of N=2 Supersymmetric SO(2r+1) Yang-Mills Theory,''
\PL {\bf B358} (1995) 273, hep-th/9504102.}
\lref\brla{A. Brandhuber and K. Landsteiner,``On the Monodromies of
N=2 Supersymmetric Yang-Mills Theory
with Gauge Group SO(2n),'' \PL {\bf 358} (1995) 73, hep-th/9507008.}
\lref\han{A. Hanany, ``On the Quantum Moduli Space of N=2 Supersymmetric
Gauge
Theories,'' \NP {\bf B466} (1996) 85, hep-th/9509176.}
\lref\arsh{P. C. Argyres and A. D. Shapere, ``The Vacuum Structure of N=2
SuperQCD with Classical Gauge
Groups,'' \NP {\bf B461}(1996) 437, hep-th/9509175.}
\lref\klmvw{A. Klemm, W. Lerche, P. Mayr, C. Vafa and N. Warner, ``Self-Dual
Strings and N=2 Supersymmetric Field Theory,''
\NP {\bf B477}
(1996) 746, hep-th/9604034.}
\lref\mawa{E. Martinec and N. Warner, ``Integrable systems and supersymmetric
gauge theory,'' \NP {\bf 459} (1996) 97, hep-th/9509161.}
\lref\kri{E. D'Hoker, I.M. Krichever and D.H. Phong, ``The Effective
Prepotential of N=2 Supersymmetric $SO(N_c)$ and $Sp(N_c)$ Gauge
Theories,'' \NP {\bf B489} (1997) 211, hep-th/9609145.}
\lref\cliff{C.V. Johnson, ``On the Orientifolding of Type II
NS-Fivebranes,'' hep-th/9705148;

S. Forste, D Ghoshal and S. Panda, ``An orientifold of the solitonic
fivebrane'', hep-th/9706057.}
\lref\gibbons{G.W Gibbons and P. Rychenkova, ``HyperKahler Quotient
Construction of BPS Monopole Moduli Spaces,'' hep-th/9608085.}
\lref\hawking{S. Hawking, ``Gravitational Instantons,'' \PL {\bf 60A}
(1977) 81.}
\lref\ptowns{P. Townsend, ``The Eleven-Dimensional Supermembrane
Revisited,'' \PL {\bf B350} (1995) 184, hep-th/9501068.}
\lref\ahha{O. Aharony and A. Hanany, ``Branes, Superpotentials and
Superconformal Fixed Points,'' hep-th/9704170.}
\lref\brha{J. H. Brodie and A. Hanany, ``Type IIA Superstrings,
Chiral Symmetry, and N=1 4D Gauge Theory Dualities,'' hep-th/9704043.}
\lref\donagiwit{R. Donagi and E. Witten, ``Supersymmetric Yang-Mills
and Integrable Systems,'' \NP {\bf B460} (1996) 299, hep-th/9510101.}
\lref\hitchin{N. Hitchin, ``The Self-Duality Equations on a Riemann
  Surface,'' {\it Proc. London Math. Soc.} {\bf 55} (1987) 59,
  ``Stable Bundles and Integrable Systems,'' {\it Duke Math. J.} {\bf
    54} (1987) 91.}
\lref\tart{R. Tartar, ``Dualities in 4D Theories with Product Gauge Groups
from Brane Configurations'', hep-th/9704198.}
\lref\athi{M. F. Atiyah and N. Hitchin, ``The Geometry and Dynamics of
Magnetic Monopoles'', Princeton University Press, 1988.}
\lref\seiprobe{N. Seiberg, ``IR Dynamics on Branes and Space-Time
Geometry'', Phys.Lett. {\bf B384} (1996) 81, hep-th/9606017.}
\lref\lll{K. Landsteiner, E. Lopez and D. A. Lowe, ``N=2 Supersymmetric
Gauge Theories, Branes and Orientifolds'', hep-th/9705199.}
\lref\sen{A. Sen, ``A Note on Enhanced Gauge Symmetries in M- and
String Theory'', hep-th/9707123.}
\lref\senO{A. Sen, ``F-theory and Orientifolds'', Nucl.Phys. B475 (1996)
562, hep-th/9605150.}
\lref\kmv{S. Katz, P. Mayr and C. Vafa, ``Exact Solution of 4D N=2
Gauge Theories - I'', hep-th/9706110.}
\lref\barb{J. L. F. Barbon, ``Rotated Branes and N=1 Duality'',
hep-th/9703051.}
\lref\haza{A. Hanany and A. Zaffaroni, ``Chiral Symmetry from Type IIA
Branes'', hep-th/9706047.}
%
%
\lref\hoo{K. Hori, H. Ooguri and Y. Oz, ``Strong Coupling Dynamics of
Four-Dimensional N=1 Gauge Theories from M Theory Fivebrane'', hep-th/9706082.}
\lref\edii{E. Witten, ``Branes and the Dynamics of QCD'', hep-th/9706109.}
\lref\biksy{A. Brandhuber, N. Itzhaki, V. Kaplunovsky, J. Sonnenschein
and S. Yankielowicz, ``Comments on the M-Theory Approach to N=1 SQCD
and Brane Dynamics'', hep-th/9706127.}
\lref\hsz{A. Hanany, M. J. Strassler and A. Zaffaroni, ``Confinement
and Strings in MQCD'', hep-th/9707244.}
\lref\nos{S. Nam, K. Oh and S.-J.Sin, ``Superpotentials of N=1
Supersymmetric Gauge Theories from M-Theory'', hep-th/9707247.}
\lref\schs{M. Schmaltz and R. Sundrum, ``N=1 Field Theory Duality from
M-Theory'', hep-th/9708015.}
\lref\fasp{A. Fayyazuddin and M. Spalinski, ``The Seiberg-Witten
Differential from M-Theory'', hep-th/9706087.}
\lref\heyi{M. Henningson and P. Yi, ``Four-Dimensional BPS-spectra via
M-Theory'', hep-th/9707251.}
\lref\mik{A. Mikhailov, ``BPS States and Minimal Surfaces'', hep-th/9708068.}
\lref\cssk{C. Cs\'aki and W. Skiba, ``Duality in Sp and SO Gauge
Groups from M-Theory'', hep-th/9708082.}
\lref\arn{V. Arnold, G Gusein-Zade and A. Varchenko, ``Singularities of
Differentiable Maps I,II'', Birkh\"auser 1985.}
\lref\threesw{N. Seiberg and E. Witten, ``Gauge Dynamics And Compactification
To Three Dimensions'', hep-th/9607163.}
%
%
\lref\sen{A. Sen, ``A Note on Enhanced Gauge Symmetries in M- and
String Theory'', hep-th/9707123.}
\lref\gM{G. Gibbons and N. Manton, Nucl. Phys. {\bf B274} (1986) 183.}
\lref\vafathree{K. Hori, H. Ooguri and C. Vafa, ``Non-Abelian Conifold
Transitions and N=4 Dualities in Three Dimensions'', hep-th/9705220.}
\lref\hw{A. Hanany and E. Witten, ``Type IIB Suprstrings, BPS Monopoles and
Three-Dimensional Gauge Theories'', Nucl.Phys. {\bf B492} (1997) 152,
hep-th/9611230.}
\lref\kkv{A. Klemm, S. Katz and C. Vafa, ``Geometric Engineering of
Quantum Field Theories'', Nucl.Phys. B497 (1997) 173, hep-th/9609239.}
\lref\jap{T. Nakatsu, K. Ohta, T. Yokono and Y. Yoshida, ``Higgs Branch
of N=2 SQCD and M theory Branes'', hep-th/9707258.}
%
%
\Title{\vbox{
\hbox{NSF-ITP-97-110}
\hbox{UCSBTH-97-19}
\hbox{\tt hep-th/9708118}
}}{\vbox{\centerline {New Curves from Branes}}}
\bigskip
\centerline{Karl Landsteiner$^{a}$ , Esperanza Lopez$^{b}$}\footnote{}{\ttsmall
$^a$ karll@cosmic1.physics.ucsb.edu, $^b$ elopez@itp.ucsb.edu}
\bigskip\centerline{\it $^{a}$Physics Department, University of California,
Santa Barbara, CA 93106, USA}
\centerline{\it $^{b}$Institute of Theoretical Physics, University of
California, Santa Barbara, CA 93106, USA}

\vskip 1cm
\centerline{\bf Abstract}

We consider configurations of Neveu-Schwarz fivebranes, Dirichlet
fourbranes and an orientifold sixplane in type $IIA$
string theory. Upon lifting the configuration to M-theory and proposing
a description of how to include the effects of the
orientifold sixplane we
derive the curves describing the Coulomb branch of $\CN=2$ gauge
theories with orthogonal and symplectic gauge groups, product gauge
groups of the form $\bigotimes_i SU(k_i)\otimes SO(N)$ and
$\bigotimes_i SU(k_i)\otimes Sp(N)$. We also propose new curves
describing theories with unitary gauge groups and matter in the
symmetric or antisymmetric representation.
\Date{\vbox{\hbox{\sl {August 1997}}}}
\goodbreak

\newsec{Introduction}

The M-theory/type IIA fivebrane has recently become a very
important tool in the study of supersymmetric four dimensional
gauge theories. The basic mechanism was first explained in \klmvw.
The fivebrane is wrapped over a Riemann surface such that the theory
living on the flat four dimensional part of the fivebrane
becomes a supersymmetric gauge theory. Which kind of gauge theory
one obtains depends on the particular Riemann surface. For $\CN=2$
theories these Riemann surfaces are precisely the ones found in
the exact solutions for the low energy effective action on the
Coulomb branch in \refs{\swi \swii \klty \arfa \haoz \aps \dasu \brla
\han \mawa
{--} \arsh}. Witten showed in \supered\ that these Riemann surfaces can
actually be derived with the help of some simple rules. He considered
configurations of fourbranes
stretched between fivebranes\foot{We follow in our
nomenclature \supered\ and will speak only of four-, five- or sixbranes
dropping the specification Neveu-Schwarz or Dirichlet since in
type IIA the dimension of the brane determines if it is NS or D.}  in
type IIA string theory and lifted this to M-theory.
In M-theory the
fourbranes are themselves fivebranes wrapped around the eleventh
dimension. What appeared in type IIA theory as a configuration
of intersecting flat four- and fivebranes becomes a single fivebrane
wrapped around a Riemann surface. Witten considered $SU(N)$ theories
and products thereof with matter in the fundamental or
bifundamental representations\foot{These theories have
recently also been derived using geometrical engineering \kkv\ in
\kmv.}. The role the Riemann surfaces play
for the brane configurations was noticed also independently in \ejs\
where an orientifold fourplane parallel to the fourbranes was added and
theories with orthogonal and symplectic gauge groups were considered.
This was combined with Witten's approach in \lll\ where
it was shown that one could derive the Riemann surfaces for a wide
variety of theories based on orthogonal and symplectic gauge groups.
The same brane configurations were investigated also in \bsty. The
Seiberg-Witten differential and the BPS spectrum in the
context of the M-theory fivebrane have been studied in \refs{\fasp \heyi {--}
\mik}.

Orientifold fourplanes parallel to the fourbranes are however not the
only possibility of obtaining orthogonal or symplectic
gauge groups. In fact it has been pointed out in the
context of brane realizations of $\CN=1$ theories already in \ejs\ \egkrs\
that one can also consider orientifold sixplanes.

In this paper we will study theories that arise by including
orientifold sixplanes into brane configurations. The basic
configuration will consist of fourbranes stretched between fivebranes
and an orientifold sixplane in between the fivebranes
such that $\CN=2$ supersymmetry in four dimensions is
preserved\foot{Although we will not consider $\CN=1$ in this paper
it should be noted that there is a rapidly growing amount of literature
and powerful new tools have also been developed for understanding
these theories from the M-theory fivebrane \ejs\ \egkrs\ \refs{\egk \barb \ejs
\brha \bsty \ahha \tart \haza \edii \hoo \biksy \hsz \nos \schs \cssk {--}
\jap}.}. There are two cases to consider. In the first one the
orientifold projects onto symplectic gauge groups on the fourbranes. In
\seiprobe\ \threesw\ it has been shown that such an orientifold
plane is described in M-theory by the Atiyah-Hitchin manifold \athi\ times
flat seven dimensional Minkowski space. It has further been
studied by Sen in \sen\ and before in a T-dual F-theory setup as
seven orientifold in \senO. In the second case the orientifold
sixplane projects onto orthogonal gauge groups on the fourbranes.
No exact M-theory description has been proposed for this case until
now. We will thus start out in section two by reviewing some
facts about orientifold projections and in section three we will
propose an M-theory description for the orientifold projecting
onto orthogonal gauge groups on the fourbranes. We show that
the Riemann surfaces for the $\CN=2$ theories with $SO$-gauge
groups follow immediately. Taking this as a convincing consistency
check for the proposed M-theory description we go on in section
four to derive the curves for theories with product gauge groups
of the form $\bigotimes_{i} SU(k_i) \otimes SO(N)$.
In section five we put a fivebrane on top of the orientifold
sixplane. Considering configurations with three fivebranes
and fourbranes stretched between them, we derive curves which we
interpret as describing theories with $SU(N)$ gauge group
and matter transforming in the symmetric representation. We test these
curves by performing several physical consistency checks.
In section six we collect some observations concerning the case with
symplectic gauge group on the fourbranes. We use these
observations to propose curves for $SU(N)$ with matter in the
antisymmetric and establish the relation to the Atiyah-Hitchin
manifold in section seven. In section eight we consider product
gauge groups of the form $\bigotimes_i SU(k_i) \otimes Sp(N)$.

\newsec{Orientifold Projections}

When Dirichlet branes come together the world-volume
theory gets promoted from a $U(1)^n$ gauge theory to $U(n)$.
Orthogonal and symplectic gauge groups can be induced by
considering Dirichlet branes in an orientifold background \joe.
Type IIA configurations of fourbranes
in the presence of an orientifold fourplane parallel to them have
been analyzed in detail in \ejs\ \egkrs\ \lll\ \bsty. We want to consider the
case of fourbranes in the background of an orientifold sixplane.

The fourbranes will lie along the directions $x_0$, $x_1$, $x_2$, $x_3$,
$x_6$ being points in $x_4$, $x_5$, $x_7$, $x_8$, $x_9$. The orientifold
sixplane will extend along the $x_0$, $x_1$, $x_2$, $x_3$, $x_7$, $x_8$,
$x_9$ directions and sit at the point $x_4=x_5=x_6=0$. It corresponds to
mod out by the space-time transformation
\eqn\O{h: \;\; (x_4,x_5,x_6)\rightarrow (-x_4,-x_5,-x_6),}
together with the world-sheet parity projection $\Omega$ and
$(-1)^{F_L}$, which changes the sign of all Ramond states
on the left \sen.
In order to obtain a single orientifold plane and avoid
charge cancellation conditions, we work in non-compact
space. Every fourbrane which does not pass through
$x_4=x_5=0$ must have a mirror image.

The described configuration breaks $1/4$ of the initial
supersymmetry, twice more than if the projection is induced by
an orientifold fourplane parallel to the fourbranes.
Let us concentrate on the massless modes of open strings with
Dirichlet boundary conditions on the fourbranes.
The $\Omega h$-eigenvalue of the massless vertex operators
$\partial_t X^{1,2,3,4}$ and $\partial_n X^{4,5}$ is $-1$. This induces
for the Chan-Paton part of the
corresponding string state $\lambda^a_{ij} |\psi,a>$ the projection
\eqn\proji{ \lambda^a = - \gamma_{\Omega h} \lambda^{aT} \gamma_{\Omega h}^{-1}
\,.}
For the vertex operator $\partial_t X^6$ and $\partial_n X^{7,8,9}$
it induces
\eqn\projii{ \lambda^a =  \gamma_{\Omega h} \lambda^{aT} \gamma_{\Omega h}^{-1}
\,.}
Depending on the choice of $\gamma_{\Omega h}$ \proji\ leaves us with
$\lambda$-matrices forming the adjoint of $SO(N)$ or the adjoint of
$Sp(N)$ \joe. Upon dimensional reduction along the $x_6$-direction we obtain
from \proji\ the bosonic fields
of an $\CN=2$ vector multiplet whereas the states from \projii\
give rise to hypermultiplets in representations other than the adjoint.

We can eliminate the matter coming from $x_6$, $x_7$, $x_8$, $x_9$
by making the fourbranes end on fivebranes that extend along
$x_0$, $x_1$, $x_2$, $x_3$, $x_4$, $x_5$ and are points in $x_6$,
$x_7$, $x_8$, $x_9$
\hw. The inclusion of such fivebranes does not break any further
supersymmetry and effectively reduces the world-volume of
the fourbranes to have only four macroscopic dimensions, i.e.
$x_0$, $x_1$, $x_2$, $x_3$.

Preserving $1/4$ of the supersymmetry allows also to include
sixbranes parallel to the orientifold sixplane.
By $T$-dualizing in $x_3$, $x_6$, a configuration of sixbranes and
fourbranes occupying directions as described is dual to a collection
of parallel sixbranes and twobranes. An orientifold plane parallel
to the sixbranes induces different projections on six- and twobranes
\gipo, therefore the same is true for the $T$-dual system of six- and
fourbranes. If we obtain an orthogonal group on the fourbranes we will
have a symplectic one on the sixbranes and vice-versa\foot{Open strings
in the 4-6 sector provide matter in the fundamental
representation for the gauge theory on the fourbranes. The sixbrane
gauge group acts then as flavor group. Let us
dimensionally reduce one direction of the fourbranes. Since the
flavor group for a four dimensional $\CN=2$ gauge theory based on an
orthogonal group is symplectic and vice-versa \arsh, this gives
another argument that the orientifold must acts with different
projections on four- and sixbranes.}. We work with the convention to
count the charges of branes and their mirror images separately. The
sixbrane charge of the
orientifold is then $+4$ if it projects onto symplectic groups on
sixbranes (orthogonal groups on fourbranes) and $-4$ if it
projects onto orthogonal groups on sixbranes (symplectic groups
on fourbranes) \joe.

\newsec{The Geometrical Set-up}

We would like to determine the non-perturbative corrections
to various brane configurations and from that derive new
Seiberg-Witten curves for $\CN=2$ supersymmetric gauge theories
in four dimensions. In \supered\ it was shown that a very efficient
way of achieving this is to lift type IIA brane configurations
to M-theory. Both four- and fivebranes of type IIA derive from a
single object in M-theory, the M-fivebrane. Their non-perturbative
description will correspond to wrap the M-fivebrane around a
certain Riemann surface. The analysis of the low energy degrees
of freedom shows that this Riemann surface is the
Seiberg-Witten curve for the effective gauge theory living on the
world-volume of the wrapped M-fivebrane.

Type IIA sixbranes correspond to Kaluza-Klein monopoles in M-theory,
or Taub-NUT spaces \ptowns. A collection of parallel sixbranes will be
given by the product of a multi-Taub-NUT space with $\IR^7$.
The multi-Taub-NUT metric \hawking\ takes the form
\eqn\multinut{
ds^2 = {V \over 4} d {\vec r}^2 + {V^{-1} \over 4} (d \tau + \vec
\omega \cdot \vec r)^2~,
}
where
\eqn\vfunc{
\eqalign{
& V= 1+ \sum_{a=1}^n {q_a \over |\vec r -\vec x_a |}~, \cr
& \vec \nabla \times \vec \omega = \vec \nabla V ~.\cr}
}
The $\vec x_a$ are the positions of the sixbranes and $q_a=1$
their charges.

Much information can be obtained however by concentrating on the
structure of the multi-Taub-NUT spaces as complex manifolds.
In one of its complex structures \multinut\ can be described
\supered\ by the equation
\eqn\An{xy=\prod_{i=1}^n (v-m_i),}
in $\IC^3$, where $n$ is the number of sixbranes present. \An\
suppresses however the positions of the sixbranes in the $x_6$
direction. Later on, when including an orientifold and using \An\ we will
always tacitly assume that the sixbranes are mirror symmetric also in
$x_6$. Recalling the brane configurations introduced in the past section
$x_4$, $x_5$, $x_6$, $x_{10}$
are the directions transversal to the sixbranes, with $x_{10}$ referring
to the eleventh dimension of M-theory. These are the directions described
by the multi-Taub-NUT space or alternatively \An. Its convenient to
identify $v=x_4+i x_5$, while $x,y$ are single-valued complex variables
associated with $x_6$ and $x_{10}$. Let us denote by $R$ the radius
of the 11-th dimension.
For large $y$ with $x$ fixed $y$ tends to $t=exp(-(x_6+i x_{10})/R)$,
while for large $x$ with $y$ fixed we have $x\sim t^{-1}$.

For brane configurations breaking $1/4$ of the initial supersymmetry
\An\ is all the necessary information from the sixbrane sector \supered.
When $m_i=0$ \An\ reduces to the equation for an $A_{n-1}$ singularity,
representing in M-theory the $U(n)$ enhancement of gauge symmetry when
the $n$ sixbranes are coincident.

In addition to sixbranes we want to consider the inclusion of an
orientifold sixplane. The M-theory metric representing an orientifold
sixplane of charge
$-4$ was shown in \threesw\ to be the Atiyah-Hitchin space \athi.
The inclusion of sixbranes was discussed recently in \sen. Far from
the orientifold the proposed metrics reduce to a multi-Taub-NUT space
containing a charge $q=-4$ at the origin and charges $q=1$ at the
positions of the sixbranes. When the sixbranes get close to the
orientifold the space develops a $D_n$ singularity \seiprobe \threesw.

We will be interested in configurations including an orientifold
sixplane of charge $+4$, which induces symplectic gauge groups on
the world-volume of sixbranes parallel to it.
In analogy with the previous cases we would
like to argue that non-perturbatively this configuration
corresponds to M-theory compactified on a 4-dimensional space
developing at some point of its moduli space a $C_n$ type singularity.
There does not exists a direct formulation of such an space.
We will assume
that the charge $+4$ orientifold affects the geometry in the same
way as the presence of $4$ coincident sixbranes.
Symplectic groups correspond to boundary singularities of $C_n$-type \arn.
We associate the boundary (or fixed point) of the $C_n$ singularity with
the point where the orientifold sits.
Thus in one of its complex structures this
space should be described by
\eqn\Osix{xy=v^4,}
for the case without sixbranes. The orientifold projection allows
only configurations of objects invariant under the $\IZ_2$
transformation \O, or its lifting to M-theory
$(x_4,x_5,x_6,x_{10})\rightarrow (-x_4,-x_5,-x_6,-x_{10})$.
In terms of the complex variables $(x,y,v)$ this
transformation reads
\eqn\or{\eqalign{& v \rightarrow -v, \cr
& x \leftrightarrow y. \cr}}
\ifig\basicfig{A configuration two fivebranes connected by parallel
fourbranes and an orientifold sixplane.}
{
\epsfxsize=2.5truein\epsfysize=2.3truein
\epsfbox{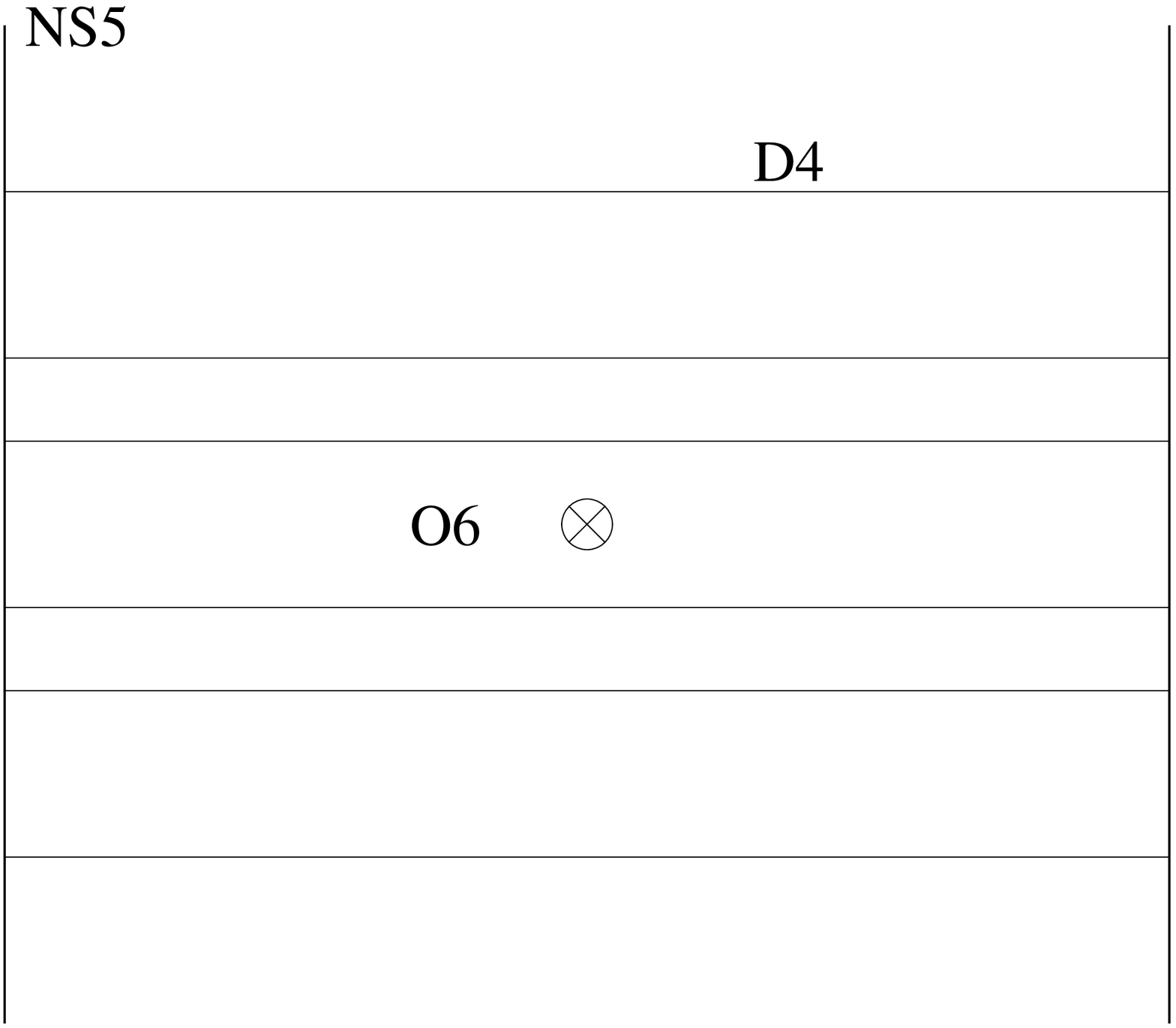}}
We consider now configurations of four- and fivebranes
invariant under \or\ in the geometry defined by \Osix.
A test of the validity of \Osix\ to describe an orientifold sixplane
of charge $+4$ will be to recover the Seiberg-Witten
curves for orthogonal groups from a configuration of
fourbranes suspended between two fivebranes as in \basicfig.

Following \supered, the exact description of
this configuration will be given in terms of a Riemann
surface $F(y,v)=0$ immersed in \Osix
\eqn\coy{F(y,v) = A(v) y^2 + B(v) y + C(v).}
We do not include semi-infinite fourbranes to the left of
the first fivebrane or to the right of the second. This implies
that $y$ or $x$ can not go to infinity for finite values of
$v$. Therefore we can set $A(v)=1$. We rewrite now \coy\
as $F'(x,v)=0$
\eqn\cox{C(v) x^2 + B(v) P(v) x + P(v)^2 =0,}
with $P=v^4$. The absence of semi-infinite fourbranes implies
that both $BP$ and $P^2$ must be divisible by $C$. At the same
time we have to impose invariance under \or, which translates
into $F'(v)=F(-v)$. These conditions have the two possible
solutions
\eqn\orstwo{\eqalign{& I: \;\;\;\;\; C(v)= v^4 \;\; , \;\;\;\;\;\;\;\;
B(v)=B(-v)  \cr
& II: \;\;\; C(v)=- v^4 \;\; , \;\;\;\;\; B(v)=- B(-v).\cr}}
Substituting option $I$ into \coy\ we obtain
\eqn\COE{y^2 + B(v^2) y + v^4 =0,}
with $B$ a generic polynomial in $v^2$ of degree $k$. Equation
\COE\ coincides with the Seiberg-Witten curves for $\CN=2$ pure
Yang-Mills with even orthogonal gauge groups $SO(2k)$.

The option $II$ implies $B(v)=v{\tilde B}(v^2)$ for ${\tilde B}$
a polynomial of degree $k$ in $v^2$. We then get
\eqn\COO{y^2 + v {\tilde B}(v^2) y - v^4 =0.}
This curve is representing a collection of $2k\! +\! 1$ fourbranes.
One of the fourbranes is stuck at $v=0$, intersecting the
orientifold, since it does not have a mirror image.
The curve \COO\ is always singular as a result of this.
By redefining $y$ (and $x$) this singularity can be
absorbed, obtaining the Seiberg-Witten curves for
$SO(2k\! +\! 1)$ gauge theories without matter.

It is worth noting
that both curves, for even and odd orthogonal gauge groups,
have singularities where branch points meet at $v=0$. They appear when
one sets the constant term in $B$ or $\tilde{B}$ to zero.
These singularities correspond to lines in the
moduli space that are present in the semi-classical regime as well as in
the quantum case. The situation
was analyzed in detail in \brla. For even orthogonal gauge groups
the corresponding singularity turned out to be an unphysical one
with trivial monodromy. For odd orthogonal gauge groups it
was shown that, although the singularity at $v=0$ exists already
in the semi-classical limit, its interpretation changes in the
quantum case, such that it can be attributed to a massless monopole. A
priori we could expect \Osix\ to describe
the orientifold only in the semi-classical regime. It is however
enough to recover the correct curves for $SO(N)$ gauge groups.
Furthermore these curves indicate the presence of the orientifold
through the singularities at $v=0$, both in the semiclassical and
fully quantum corrected regimes. This shows that \Osix\ is valid
even in the strong coupling regime.

The curves for orthogonal groups with fundamental matter can
be derived by including semi-infinite fourbranes or by adding
sixbranes parallel to the orientifold sixplane. We will treat
the inclusion of sixbranes in the next section, when
considering more general brane configurations.

\newsec{Product Groups: $\prod_{i} SU(k_i) \times SO(k)$}

\ifig\product{A configuration fivebranes connected by parallel
fourbranes, sixbranes and an orientifold sixplane.}
{
\epsfxsize=3.0truein\epsfysize=2.5truein
\epsfbox{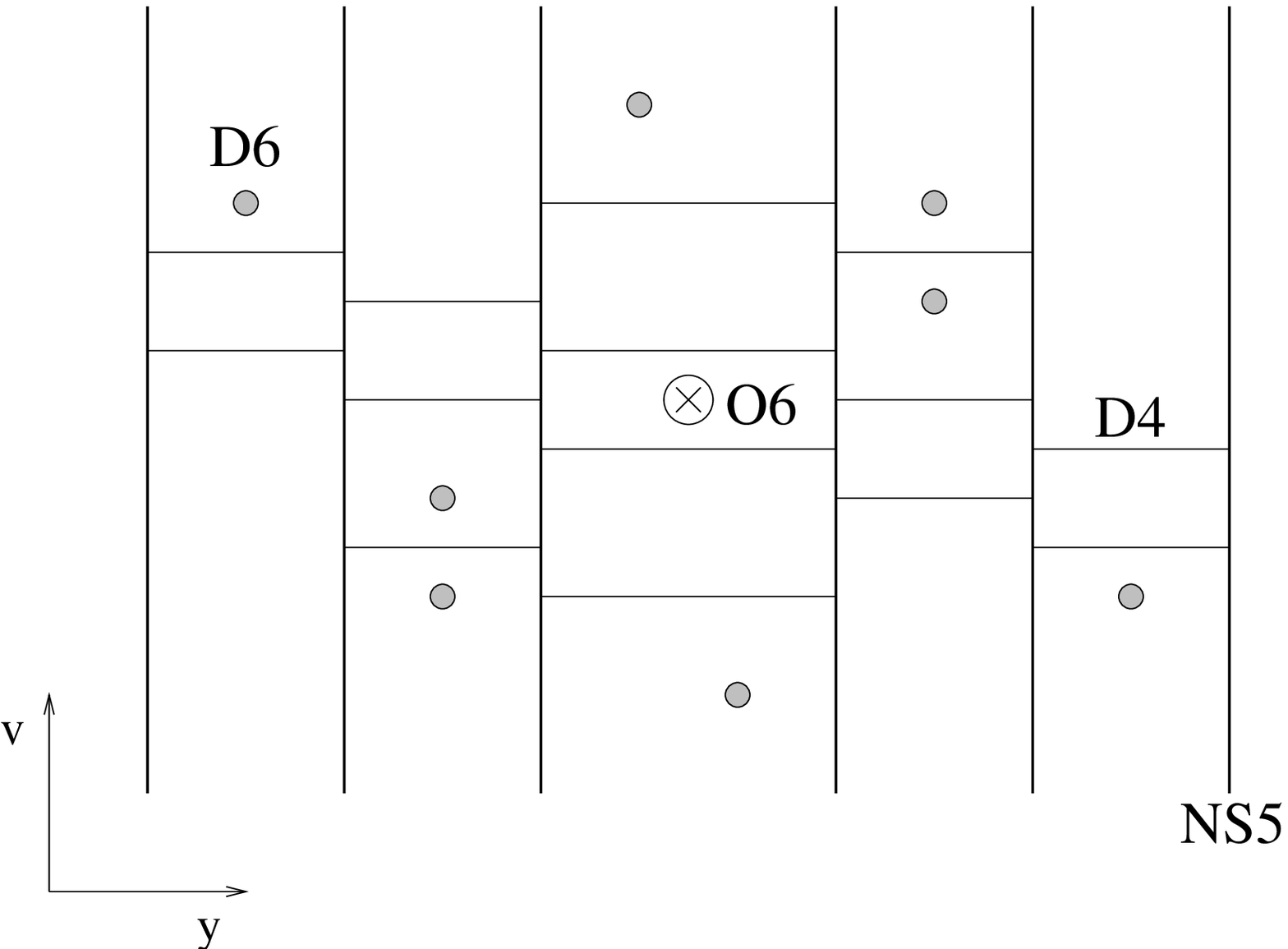}}
Let us analyze a configuration of $2n$ fivebranes
with $k_i$ fourbranes suspended between the $i$-th and $i+1$-th
fivebranes in the presence of the orientifold sixplane. We also
allow for $d_i$ sixbranes placed between the $i$-th and $i+1$-th
fivebranes. This collection of branes must be arranged in a
symmetric way under the space-time projection induced by
the orientifold (see \product), therefore
\eqn\kd{k_i=k_{2n-i} \;\;, \;\;\;\;\;\;\; d_i=d_{2n-i}.}

The fluctuations of the $k_i$ fourbranes for $i=1,\ldots n-1$ are not
constraint by the orientifold but merely related to the mirror images.
Therefore on the world-volume of the fourbranes this configuration
will induce the product gauge group $\prod_{i=1}^{n-1} SU(k_i) \times
SO(k_n)$\foot{We get special unitary groups because the overall $U(1)$
factor is frozen out by the finite energy condition on the fivebranes
as explained in \supered.}.
The matter content consists of $n-2$ hypermultiplets transforming
in the bifundamental representation $({\bf k_i},{\bf {\bar k}_{i+1}})$
of the $SU$ factors, for $i=1,..,n-2$, and one hypermultiplet in the
$({\bf k_{n-1}},{\bf k_n})$,
with ${\bf k_n}$ denoting now the vector representation of $SO(k_n)$.
In addition there are $d_i$
fundamental hypermultiplets for each factor $SU(k_i)$, and
$d_n/2$ fundamentals for $SO(k_n)$.

We want to derive the Riemann surface encoding the exact
solution for the described $\CN=2$ gauge theory.
In order to take into account the effect of the sixbranes
we generalize \Osix\ to
\eqn\Od{xy=P(v)=(-1)^d v^4 \prod_{i=1}^{2d} (v - m_{i}),}
with $2d=\sum d_i$ and for each mass parameter $m_i \neq 0$
there exits $m_j=-m_i$. We have introduced the factor $(-1)^d$
for convenience. We now define
\eqn\J{J_i = A_i \prod_{j=l_{i-1}+1}^{l_i} (v-m_j)}
with $l_i=\sum_{j=1}^i d_j$ and
\eqn\as{\eqalign{&  A_i =1 \;\;\;\;\;\;\;\;\;\;\;\;\;\;\;\; i=1,..,n-1 \cr
&  A_n = (-1)^{d_n/2} \; v^4 \cr
& A_i =(-1)^{d_i}  \;\;\;\;\;\;\;\; i=n+1,..,2n-1 .\cr}}
The polynomials $J_i$ represent the contribution to the space
\Od\ from the sixbrane charge sources, i.e. sixbranes and
orientifold sixplane, between the $i$-th and $i+1$-th fivebranes.
They satisfy $P=\prod_{i=1}^{2n-1} J_i$. The $\IZ_2$ invariance
with respect to the orientifold implies $ m^{(2n-i)}_j =
- m^{(i)}_j $. Using this, the polynomials $J_i$ verify
\eqn\js{J_i(v)=J_{2n-i}(-v).}

In \supered\ the Riemann surface representing a collection of
fivebranes with fourbranes suspended between them
and immersed in an space of the form \Od\ was determined. When no semi-infinite
fourbranes are present we have
\eqn\sup{y^{2n} + g_1(v) y^{2n-1} + \dots + g_i(v) \prod_{s=1}^{i-1}
J^{i-s}_s \; y^{2n-i} + \dots + f \prod_{s=1}^{2n-1} J^{2n-s}_s =0,}
with $f$ a constant and $g_i$ polynomials in $v$ of degree
$k_i$ given by the number of fourbranes suspended between the $i$-th
and $i+1$-th fivebranes
\eqn\g{g_i(v)= c_{0,i} v^{k_i} + c_{1,i} v^{k_i-1}
+ c_{2,i} v^{k_i-2} + \dots + c_{k_i,i}.}

In our case we have in addition to impose
symmetry under $(v,y,x) \rightarrow (-v,x,y)$. This,
together with the property \js\ of the polynomials $J_i$, allows
for the following two sets of solutions
\eqn\ors{\eqalign{& I: \;\;\;\;\; f=1 \;\; , \;\;\;\;\;\;\;\;\;\;
g_i(v) = g_{2n-i}(-v) \cr
& II: \;\;\; f=-1 \;\; , \;\;\;\;\;\;\;
g_i(v) = - g_{2n-i}(-v) . \cr}}
These conditions leave the polynomials $g_i(v)$ for
$i=1,..,n-1$ unconstrained. This is necessary in order they represent
$SU$ group factors. The solution $I$ implies $g_n(v)=g_n(-v)$
corresponding to product gauge groups of unitary groups
with an even orthogonal group, $\prod_{i=1}^{n-1} SU(k_i) \times
SO(2k)$. The solution $II$ implies $g_n(v)=-g_n(-v)$ and
is associated with chains including an odd orthogonal group,
$\prod_{i=1}^{n-1} SU(k_i) \times SO(2k+1)$.

The coefficients $c_{\alpha,i}$, $\alpha \ge 2$, in \g\
describe the Coulomb branch of the obtained gauge theory.
Equations \ors\ reduce the number of independent $c_{1,i}$
to $n-1$, which represent bare masses for the $n-1$
hypermultiplets in the bifundamental representations.
The coefficients $c_{0,i}$ are associated with ratios
of gauge couplings. Rescaling $v$ and $y$ we obtain $n-1$
independent $c_{0,i}$ as corresponds to a product gauge
group with $n$ factors\foot{We can not restrict to $n-2$
independent $c_{0,i}$ by rescaling separately
$v$ and $y$ since this will introduce an additional
parameter in \Od.}.

$\CN=2$ gauge theories based on orthogonal groups with matter
content in the vector representation have symplectic flavor
symmetry groups \arsh. Let us consider our brane
configuration in the absence of sixbranes. The matter
content for the orthogonal factor group $SO(k_n)$ consists
of $k_{n-1}$ $\CN=2$ hypermultiplets in the vector
representation. The associated flavor group is $Sp(2k_{n-1})$.
In the brane configuration only a $SU(k_{n-1})$ subgroup is
manifest, which is given by the next factor group in the
chain.

\newsec{$SU(N)$ with matter in the symmetric}

Until now we always placed the orientifold in between two
fivebranes. This is however not the only possibility.
In this section we will identify and study the theory that arises
when we place a fivebrane on top of the orientifold sixplane.
Such a configuration has been studied from the conformal
field theory point of view recently in \cliff.
More precisely we place one fivebrane on top of the orientifold
sixplane and another fivebrane to the right of it. Further there
be $N$ fourbranes stretching in between the fivebranes. To the left
of the orientifold sixplane we have of course the mirror image of
this brane configuration. We do not include sixbranes in this
configuration. Again we assume that the orientifold can be
described by \Osix.
The polynomial describing the wrapped M-theory fivebrane will now
be of third order in $y$. For brane configurations
with an odd number of fivebranes the options $I$ and $II$ in \ors\
are equivalent, but just related by $y \rightarrow -y$.
Thus we obtain
\eqn\symcurve{ y^3 + y^2 \prod_{i=1}^N (v-a_i) + (-)^N y v^2
\prod_{i=1}^N (v+a_i) + v^6 =0 \, .}
One can readily check that this curve is symmetric under the
simultaneous replacements $v \rightarrow -v$ and $y \rightarrow v^4/x$.

In order to identify the gauge theory whose Coulomb branch is described
by \symcurve, we note that the gauge theory will be $SU(N)$. This is
because the orientifold does not constrain the fluctuations of the
fourbranes but relates them to their mirror images on the other side of
the middle fivebrane. The endpoints
of fourbranes to the right of the orientifold sixplane will
give rise to hypermultiplets. If we pretend for a moment that the
two fourbrane groups were independent of each other, we would obtain
a product gauge group with matter transforming in the fundamental
with respect to each factor. Since in our case both gauge groups
factors have to be identified, the bifundamental turns into a two index
representation of $SU(N)$. It can not be the adjoint since not all
degrees of freedom of the
adjoint are really available. A string stretching from one
fourbrane on the right to a fourbrane on the left comes inevitably
with its mirror image. Thus we conclude that the gauge theory on the
M-theory fivebrane will contain a hypermultiplet transforming
in either the symmetric or antisymmetric representation of $SU(N)$.
This is possible because hypermultiplets contain simultaneously
fields transforming under the fundamental and anti-fundamental of
$SU(N)$.

We can use the symmetry properties of \symcurve\ in order to decide
which case is realized. It is well known that the curves describing
the Coulomb branch of $\CN=2$ gauge theories have to reflect the breaking
of the $U(1)_R$ symmetry to $\IZ_{b_0}$, with $b_0$ being the one-loop
beta function coefficient. This $U(1)_R$ symmetry appears as
rotations in the $v$-plane. It is spontaneously broken if the
fourbranes are not all placed at the origin.
Let us assign now $R$-charge $1$ to $v$ and
$a_i$ and R-charge $N$ to $y$. If we perform now such a
transformation with $x\rightarrow \exp(i \a)$ and $y\rightarrow \exp(i
N \a)$ we find that the curve \symcurve\ stays invariant if we chose $\a
= 2\pi {n \over N-2}$, showing a $\IZ_{N-2}$ symmetry. The one loop beta
function coefficient of the gauge theory is thus $b_0 = N-2$. This has
to equal $b_0 = 2N - 2 I_m$ where $I_m$ is the index of the
representation of the hypermultiplet. We find thus $I_m = (N+2)/2$
which is the index of the symmetric representation of $SU(N)$.

In the remainder of the section we will concentrate in performing
several consistency checks of the proposed curves. In order to
do so it is convenient to introduce the scale $\Lambda$ of the gauge
theory. The space \Osix\ reads then $xy= \Lambda^{2N-4} v^4$.
We further observe that \symcurve\ is singular at $v=y=0$ since the middle
fivebrane passes through the orientifold. We
can remove this singularity by redefining $y \rightarrow v^2 y$. The
precise statement is now as follows. The subspace of the Jacobian
selected by the cycles invariant under $(v,y) \rightarrow
(-v,\Lambda^{2N-4}/y)$ of the curve
\eqn\newsym{ v^2 y^3 + y^2 \prod_{i=1}^N (v - a_i) + (-)^N
\Lambda^{N-2} y \prod_{i=1}^N (v+a_i) + \Lambda^{3N-6} v^2 =0\,.}
parameterizes
the Coulomb branch of the moduli space of $\CN=2$ gauge theories with a
hypermultiplet in the symmetric representation. That the invariant
subspace of the Jacobian is the relevant one can be seen be noting that
the Seiberg-Witten differential $\lambda_{SW} = v {dy\over y} $ stays
invariant. The $N$ parameters $a_i$ represent the positions of the
fourbranes in the classical brane configuration. The distance between the
average position of the fourbranes on the left and the average
position of the fourbranes on the right is the mass of the
hypermultiplet in the symmetric
\eqn\massym{ m = {2\over N} \sum_{i=1}^N a_i\, .}
Thus we see that the curve has just the correct number of free
parameters.

A rather nontrivial consistency check for the curves \newsym\ is a
follows. On the classical Coulomb moduli space there are
submanifolds where the $SU(N)$ group is only partially broken
and factorizes as $SU(N) \rightarrow \bigotimes_{i=1}^l SU(n_i)\otimes
U(1)^{l-1}$. The hypermultiplet in the symmetric will
split up in various matter fields transforming under $SU(n_i)$ factors
according to the decomposition of the symmetric representation. More
precisely, if we parameterize the Coulomb branch by $\Phi$ being a
matrix in the fundamental of $SU(N)$ then the above breaking pattern
corresponds to choosing $\Phi =
diag(e_1,\cdots,e_1;e_2,\cdots,e_2;\cdots;e_l,\cdots,e_l)$. Here each
$e_i$ appears $n_i$ times, $e_i\neq e_j$, and $\sum n_i e_i = 0$. The
weights of the symmetric representation can be obtained as the
symmetric product of the weights of the fundamental of $SU(N)$.  We
denote the later by $\lambda_I$, $I=1,\cdots,N$. Under the symmetry
breaking the fundamental of $SU(N)$ decomposes
into the fundamentals of the $SU(n_i)$ factors whose weights we denote
by $\lambda_i$. The decomposition of the symmetric is thus given by
\eqn\decompsym{ \lambda_I \bigotimes_{sym} \lambda_J = \bigoplus_{i\leq
j} \lambda_i  \bigotimes_{sym} \lambda_j \, .}
The masses of the various matter fields are given by $m_i = 2 e_i$ for
a hypermultiplet in the symmetric of $SU(n_i)$ and $m_{ij} = e_i + e_j$
for a hypermultiplet in the bifundamental of $SU(n_i)\otimes SU(n_j)$.
We can make these matter fields massless by choosing $m$ to cancel
either $m_i$ or $m_{ij}$. The light degrees  of freedom are then the
vector multiplets of the gauge group factors and a hypermultiplet in the
symmetric of one $SU(n_i)$ or in the bifundamental of $SU(n_i)\otimes
SU(n_j)$. This symmetry breaking pattern has to be respected by the
curve \newsym. If we take the symmetry breaking scale to be large, then
the curve should reduce to
the curves describing the gauge groups factors with the corresponding
matter content after performing an appropriate scaling limit.

Let us investigate these scaling limits in detail now. We start with scaling
to a smaller gauge group $SU(n_i)$ with matter in the symmetric. This can be
achieved by choosing $e_i = E$, $E$ large, and $m=-2E$. Taking
into account \massym\ this can be translated into the positions of the
fourbranes $a_i=e_i+{m \over 2}$. We also allow for small fluctuations
$\delta a_k$ in the
positions of the fourbranes corresponding to the $SU(n_i)$ factor but
suppress these fluctuations for the rest of the fourbranes. The curve takes
the form
\eqn\facti{\eqalign{&v^2 y^3 + y^2 \prod_{j\neq i}(v+E-e_j)^{n_j}
\prod_{k=1}^{n_i}(v-
\delta a_k) + \cr
&(-)^N \Lambda^{N-2} y \prod_{j\neq i}(v-E+e_j)^{n_j}
\prod_{k=1}^{n_i}(v+\delta a_k) + \Lambda^{3N-6} v^2 = 0\, .}}
After rescaling $y \rightarrow E^{N-n_i} y$ this becomes
\eqn\factii{\eqalign{&v^2 y^3 + y^2 \prod_{j\neq i}({v\over E}+1-
{e_j\over E})^{n_j} \prod_{k=1}^{n_i}(v-
\delta a_k) + \cr
& (-)^N E^{-N+n_i} \Lambda^{N-2} y \prod_{j\neq i}({v\over E}-1+
{e_j\over E})^{n_j}
\prod_{k=1}^{n_i}(v+\delta a_k) + E^{-3N+3n_i}\Lambda^{3N-6} v^2=0\, .}}
Now we can perform a renormalization group matching
\eqn\rgmatchi{\left( {\Lambda \over E} \right)^{N-2} = \left(
{\tilde{\Lambda}
\over E} \right)^{n_i-2}\,.}
Taking $E$ to infinity but keeping $\tilde{\Lambda}$ fixed \factii\
reproduces \newsym\ with $N=n_i$.

Alternatively we could have chosen $m=0$. The light degrees of freedom consist
then only of the vector multiplets of the gauge group factors. We concentrate
now on the $i$-th factor. If we introduce
$\xi = v-e_i$, rescale
$y\rightarrow E^{N-n_i-2} y$ and suppress fourbrane fluctuations outside of
this factor. We further assume that $e_j\neq - e_i$ since then we have
additional massless matter in the bifundamental of $SU(k_i)\otimes SU(k_j)$.
The curve can be written as
\eqn\factiii{\eqalign{&({\xi \over E}+1)^2 y^3 + y^2 \prod_{k=1}^{n_i} (\xi-
\delta a_k)
\prod_{j\neq i} ({\xi \over E}+1-{e_j\over E})^{n_j} +\cr
&(-)^N \Lambda^{N-2} y E^{-N+2+2n_i} \prod_{k=1}^{n_i} ({\xi\over E} + 2
+ {\delta a_k \over E})^{n_j} \prod_{j\neq i} ({\xi\over E} + 1 +
{e_j\over E})^{n_j}
+ \Lambda^{3N-6} E^{-3N+3n_i +6}=0.}}
After a RG-matching
\eqn\rgmatchii{\left( {\Lambda \over E} \right)^{N-2} = \left(
{\tilde{\Lambda}
\over E} \right)^{2n_i}\,,}
and taking $E$ to infinity this takes the form of the curve for pure $SU(n_i)$
gauge theory.

Next we chose $m=-e_i-e_j$. We expect to find the curve for $SU(n_i)\otimes
SU(n_j)$. We set now ${e_j-e_i \over 2}=E$. The curve takes the form
\eqn\factiv{\eqalign{& v^2 y^3 + y^2 \prod_{k=1}^{n_i}(v+E -
\delta a_k)  \prod_{l=1}^{n_j}(v-E -
\delta b_l) \prod_{p\neq i,j}(v-a_p)^{n_p} + \cr
&(-)^N \Lambda^{N-2} y \prod_{k=1}^{n_i}(v-E+
\delta a_k)  \prod_{l=1}^{n_j}(v+E+\delta b_l)
\prod_{p\neq i,j}(v+a_p)^{n_p} + \Lambda^{3N-6}v^2 = 0 \,.}}
Since below the scale $E$ we have two gauge groups factors the RG-matching
has to fulfill now
\eqn\rgmatchiii{\left( {\Lambda \over E} \right)^{N-2} = \left(
{\tilde{\Lambda}_i
\over E} \right)^{2n_i-n_j} = \left(
{\tilde{\Lambda}_j
\over E} \right)^{2n_j-n_i}\,.}
 From this we see that as long as $n_i\neq n_j$,  in the limit
$E\rightarrow \infty$ one of the two gauge
group factors inevitably ends up at infinite or vanishing coupling.
In order to avoid this we set $n_i=n_j=n$. Upon introducing $\xi = v+E$,
rescaling
$y\rightarrow E^{N-n-2} y$ and absorbing some numerical factors into
$\tilde{\Lambda}$ \factiv\ becomes
\eqn\bifund{y^3 + y^2 \prod_{k=1}^{n} (\xi - \delta a_k) + \tilde{\Lambda}^n
y \prod_{l=1}^n (\xi - \delta b_l) + \tilde{\Lambda}^{3n} =0 \,.}
This is the correct form of the curve for $SU(n)\otimes SU(n)$ with matter
transforming in the bifundamental \supered\ \kmv.

As a final example for scaling limits of \newsym\ let us take the mass $m$ to
infinity. It can be seen easily that in this limit, and after introducing
$\xi = v-{m \over 2}$, $y \rightarrow m^{-2} y$ and the RG-matching
$\Lambda^{N-2} m^{N+2} = \tilde{\Lambda}^{2N}$, the curve
\eqn\factvi{v^2 y^3 + y^2 \prod_{i=1}^N ( v - {m\over 2}  - e_i ) +
(-)^N \Lambda^{N-2} y \prod_{i=1}^N (v+{m\over 2} + e_i) + \Lambda^{3N-6}v^2
=0\,,}
reproduces the one for the $SU(N)$ theory without matter.

The case of $N=2$ is special. First we can not eliminate all possible numerical
factors by scalings of $y$ or $v$. Moreover the symmetric of $SU(2)$ is
actually the adjoint. Therefore the curve
\eqn\softly{ v^2 y^3 + y^2 (v^2 - m v - u) + e y (v^2 +m v-u)
+ e^3 v^2 =0 \,,}
should describe the softly broken $\CN=4$ $SU(2)$ theory. That we can not
eliminate the dimensionless complex number $e$ by  rescalings is then the
expected behavior. It determines the gauge coupling of the softly broken
$\CN=4$ theory.
The discriminant of \softly\ is given by
\eqn\disci{\Delta = e (1+e) m u \left(m^4 + 8 m^2 u + 40 em^2 u + 16
u^2 - 96 e u^2 + 144 e^2 u^2\right)\,.}
Since the curve \softly\ is at most quadratic in $v$, it is
hyperelliptic. Its genus is two. If we write it as $v^2 A + v B +C = 0$
and introduce the new coordinate
$\tilde{v} =  {v-B \over 2A}$, the curve takes the form
\eqn\softlyhyp{\tilde{v}^2 = y^2 m^2(y-e)^2
+ 4 (y+e)^2 (y^2+y(1-e) + e^2) u y \, .}
At $m=0$ after absorbing the overall $y+e$ factor into $\tilde{v}$ we
are left with a torus. The complex structure of this torus depends only
on $e$ and its degenerations are located at $\Delta_e = e
(1+e)(1-3e)=0$. Notice that there is a new factor $(1-3e)$ that did not
appear in \disci. We can make its appearance manifest by redefining $m
= (1-3e) \tilde{m}$. Remember that in the case of
theories with vanishing beta function one can reparameterize
the moduli space with the help of arbitrary functions of the
dimensionless coupling constant with the only condition that the new
moduli coincide with the classical ones in the weak coupling limit \aps\
\arsh.
In our case this sets the weak coupling limit at $e=0$.
That this is so can be deduced from the double
scaling limit to obtain pure $SU(2)$, which works with $e\rightarrow 0$ and
$m\rightarrow \infty$ as in \factvi. The
discriminant \disci\ is given now by
$\Delta = \tilde{m} \, . \Delta_e .\Delta_u$ where we set
\eqn\discu{\Delta_u = u \left( \tilde{m}^4 (1-3e)^2 + 8\tilde{m}^2
u + 40 e \tilde{m}^2 u + 16 u^2\right)\,.}
The singularities in the $u$-plane a located at $u=0$ and $u=u_\pm$ with
\eqn\upm{u_\pm = \left[- (1+5e)/4 \pm \sqrt{e(1+e)}\right] \tilde{m}^2 \,.}

We want to compare this now with the singularities of the curve for
the softly broken $\CN=4$ theory \swii\ \donagiwit
\eqn\swcurve{y^2 = \prod_{i=1}^3 (x-e_i U - {1\over 4} e_i^2 M^2)\,.}
For $M=0$ the $e_i(\tau)$ are the roots of a cubic describing a torus of
modular parameter $\tau$. Their relation to Jacobi theta functions is
given by $e_1-e_2 = \theta_3(\tau)^4$,  $e_3-e_2 = \theta_1(\tau)^4$
and  $e_1-e_3 = \theta_2(\tau)^4$.
Its singularities are located at $\Delta_U . \Delta_\tau =0$ with
\eqn\deltatau{ \Delta_\tau = \prod_{i\le j} (e_i-e_j)\, ,}
and
\eqn\deltaw{\Delta_U = \prod_{i=1}^3 (U - e_i (M/2)^2) \,.}
If we shift $U \rightarrow U + e_1 (M/2)^2$ and identify $U$ with $u$
we obtain the equations
\eqn\taue{\eqalign{&e_2-e_1 = \left[-(1+5e) + 4\sqrt{e(1+e)}\right]
(\tilde{m}/M)^2 \, \cr
& e_3-e_1 = \left[-(1+5e) - 4\sqrt{e(1+e)}\right] (\tilde{m}/M)^2 \,.}}
After a little algebra we extract from that
\eqn\defe{ e = { -(5h^2-8) \pm 8 \sqrt{1-h^2} \over 25h^2-16} \, ,}
where
\eqn\hdef{h = {e_2-e_3 \over 3 e_1} \,.}
The zeroes of $\Delta_\tau$ are at $\tau = (i\infty, 0, 1)$.
These values correspond to $e_2=e_3$, $e_1=e_3$ and $e_1=e_2$.
Thus at weak coupling $\tau = i \infty$ we have $h=0$ and $e=0$ or
$e=-1$. The values $\tau = 0$ and $\tau = 1$ are mapped onto $h=1$ and
correspond to $e=1/3$. These values are the zeroes
of $\Delta_e$. It is a nontrivial check of consistency that
after matching of $\Delta_U$ to $\Delta_u$ we obtain automatically
the matching between $\Delta_\tau$ and $\Delta_e$.

The discriminant \disci\ contains however also a singularity at
$\tilde{m} =0$. We can study it by looking at the holomorphic
differentials on the curve \softlyhyp. Since it has genus two $\omega_1
= {dy \over \tilde{v}}$ and $\omega_2 = {y dy \over \tilde{v}}$ form a
basis of holomorphic differentials. The orientifold $\IZ_2$ involution
acts as $(\tilde{v},y) \rightarrow
(-e^3/y^3 \tilde{v}, e^2/y)$ in the new coordinates. It also
acts as $(A,C) \rightarrow e^3/y^3 (A,C)$ and
$B \rightarrow - e^3/y^3 B$, where $A,B$ and $C$ as defined above. On
the holomorphic differentials
it acts as $e \omega_1 \leftrightarrow \omega_2$. $\Omega_\pm = e
\omega_1 \pm\omega_2$ form a basis of even and odd holomorphic
differentials. As needed we find $\partial_u \lambda_{sw} = \Omega_+$.
When $\tilde{m}$ goes to zero the genus of the curve
drops. Then we can absorb the overall factor of $(y+e)$ into the
definition of $\tilde{v}$. Doing this
we notice that $\Omega_+$ descends to the unique holomorphic
differential on the remaining torus! Since the physically relevant
periods stem only from $\Omega_+$ none of them vanishes at this
particular degeneration. Thus we do not expect any new massless
state at $\tilde{m} = 0$.

We think that the above arguments are rather strong evidence in
favor of the conjecture that \softly\ describes the softly broken
$\CN=4$ theory correctly.

\newsec{$SU(N)$ with matter in the antisymmetric}

In the previous sections we have analyzed brane configurations
in the background defined by an orientifold sixplane of charge
$+4$. Now we want to consider situations including an
orientifold sixplane of the opposite charge, i.e. $-4$.
Some configurations in the presence of a negative charge
orientifold sixplane, or its $T$-dual orientifold sevenplane, have
been studied in \senO\ \seiprobe\ \threesw.

In the absence of branes that probe the geometry induced by the
orientifold, the sign of the orientifold charge is a matter of
convention. In \threesw\ it was shown that the geometry seen
by a twobrane probe parallel to an orientifold sixplane which projects
onto symplectic groups on the twobrane,
is given by the Atiyah-Hitchin space \athi. Far from the
orientifold this space reduces to a multi-Taub-NUT space
containing a charge $q=-4$ at the origin \gM\ \threesw\ \sen.
An approximate description of this space as a complex
manifold, valid far from the orientifold, will be given by
\eqn\oo{xy= \Lambda^{2N+4} v^{-4} .}

We consider now a configuration of two fivebranes and fourbranes
in the space defined by \oo. As before, the presence of the orientifold
forces any collection of objects to be arranged in a symmetric way under
the $\IZ_2$ transformation \or\ (see \basicfig).
Imposing invariance under $(v,y,x) \rightarrow
(-v,x,y)$ and following the same steps as in section 3, we obtain the
following curve associated to our configuration
\eqn\symp{y^2 + y \, \left( b(v^2)+A v^{-2} \right) +
\Lambda^{2N+4} v^{-4} =0.}
We have reduced ourselves to the option $I$ in \ors. This constrains
the number of fourbranes $N$ to be even.
We notice two different contributions to \symp, $b(v^2)$ and $A v^{-2}$.
$b$ is a polynomial in $v^2$ of degree $k=N/2$ whose coefficients,
according to \supered,
we associate with Casimirs of a gauge theory. $A$ is an
undetermined constant. The additional term $A v^{-2}$ is allowed
due to the $v^{-4}$ contribution of the orientifold. For large
values of $v$ this term is irrelevant, however it dominates
for small $v$.

We interpret the extra term $A v^{-2}$ as a manifestation that close
to $v=0$ the space \oo\ does not provide a good description
of the orientifold background. It is missing strong coupling effects
that would modify the geometry into that of the
Atiyah-Hitchin space. However when formulating curves in \oo\ the
possibility to take into account strong coupling effects reappears.
It is encoded in the presence of additional monomials in negative
powers of $v$, as it is the case of $A v^{-2}$ in \symp.
Indeed, by setting $A=2 \Lambda^{N+2}$ and rescaling $y \rightarrow v^{-2} y$
we recover the Seiberg-Witten curves for $\CN=2$ pure Yang-Mills theory with
gauge group $Sp(2k)$ \mawa\ \arsh\ \kri
\eqn\Symp{y^2 + y \, \left( v^2 b(v^2)+ 2 \Lambda^{N+2} \right)
+ \Lambda^{2N+4} =0.}
We observe the direct dependence of the constant $A$
of strong coupling effects whose characteristic
scale is set by $\Lambda$.

A similar situation was encountered \lll\ when analyzing brane
configurations in the presence of an orientifold fourplane. For
configurations providing symplectic gauge groups, strong coupling
effects associated with the orientifold led to the additional
factor encountered in \symp. In that case the analysis of the
theory on the world-volume of the fivebranes allowed to fix the
constant $A$ and recover the symplectic curves. In the present
case we do not know how to carry out such an analysis. In order
to fix this ambiguity we use as guideline the matching with
expected results from field theories.

Let us analyze next a configuration containing three fivebranes.
Invariance under \or\ leads to the curve
\eqn\antisym{y^3 + y^2 \, \left(p(v) +B v^{-1}  + A v^{-2} \right) +
\Lambda^{N+2}v^{-2}  y
\, \left( q(v) - B v^{-1} + A v^{-2} \right) +  \Lambda^{3N+6} v^{-6} =0,}
where $p(v)=q(-v)=\prod_{i=1}^N (v-a_i)$ and $N$ can be
even or odd now. In the present case two
additional factors in negative powers of $v$ are allowed,
$A v^{-2}$ and $B v^{-1}$ with $A,B$ constants. Using the
same reasoning as in the beginning of section 5, we conclude
that this curve should describe an $SU(N)$ theory with matter
in a tensor representation. The natural candidate for the matter
representation in the present case is the antisymmetric.

In order to check this we analyze the breaking of the $U(1)_R$
symmetry to a residual $\IZ_{b_0}$. We assign $R$-charge $1$ to
$v$ and $a_i$, and $N$ to
$y$. Under a $U(1)_R$ transformation the curve \antisym\ remains
invariant only if the angle verifies $\alpha=2\pi {n \over N+2}$,
showing a $\IZ_{N+2}$ symmetry. Since for $\CN=2$ theories this
residual symmetry is directly related to the one-loop beta function
coefficient we have
\eqn\betaA{b_0=2N - 2I_m = N+2,}
where $I_m$ is the index of the matter representation. Therefore
we obtain $I_m=(N-2)/2$ which is the index of the antisymmetric
representation of $SU(N)$.

In the previous argument we have ignored the terms associated to the
constants $A$ and $B$. Equation \antisym\ can be made $U(1)_R$
invariant if we assign convenient $R$-charges to $\Lambda$, $A$
and $B$. The parameter $\Lambda$ must be assigned charge $1$.
The $R$-charge of $A$ and $B$ has to be $N+2$ and $N+1$ respectively.
We have argued
that $A$ and $B$ are associated with strong coupling effects. Based on
this we set $A \sim \Lambda$ and $B \sim \Lambda /m$ with $m$
the mass of the antisymmetric hypermultiplet. Such a $B$ coefficient
will introduce large effects at $m \rightarrow 0$ even for
small $\Lambda$. Since no such effects are expected in gauge
theories we set $B=0$.

To fix the precise value of the constant $A$ we consider the
case in which $p(v)=q(v)$. This can be achieved only for $N$
even, by turning off the coefficients in $p$ that multiply
odd powers of $v$. The curve \antisym\ then factorizes in
two pieces
\eqn\antifact{\left( y+ \Lambda^{N+2} v^{-2} \right)
\left( y^2 + y \, \left( p(v^2)+ (A-\Lambda^{N+2})v^{-2} \right) +
\Lambda^{2N+4} v^{-4} \right)=0.}
This corresponds to fourbranes on the left and right of the
middle fivebrane recombining into single fourbranes. In this
situation the middle fivebrane decouples, can be pulled
away, and it is just described by the first factor in \antifact.
The second factor will now describe an even number of fourbranes
suspended between two fivebranes, and therefore must reproduce
the Seiberg-Witten curve for symplectic gauge groups without
matter. This condition sets $A=3 \Lambda^{N+2}$. We will assume
that the same value of $A$ holds for $N$ odd.

Rescaling $y \rightarrow y v^{-2}$, we
propose finally the following curves for representing $SU(N)$
with a hypermultiplet transforming in the antisymmetric
representation
\eqn\AS{y^3 + y^2 \left(v^2 \prod_{i=1}^N (v-a_i) + 3 \Lambda^{N+2}\right)
+ \Lambda^{N+2} y \left( (-)^N v^2 \prod_{i=1}^N (v+a_i) +3\Lambda^{N+2}
\right)  + \Lambda^{3N+6}=0.}
These curves have the correct behavior under analogous scaling
limits to that studied in section 5. We refer to that
section for details. At $v=0$ \AS\ has a triple solution in $y$.
Configurations containing two fivebranes satisfy a similar property,
the curve \Symp\ has a double solution in $y$ at $v=0$.
As it is the case for the symplectic curves, the point $v=0$
is a singularity of \AS\ for any value of the Casimirs
and mass parameter. In the next section we will rewrite \AS\ in a
form that eliminates the singularity at $v=0$.
We will now analyze in detail the cases $N=2,3$ as consistency
checks of our curves.

For $N=2$ the previous curve reads
\eqn\AStwo{y^3 + y^2 \left(v^2 (v^2 + vm +u) + 3 \Lambda^4 \right)  +
\Lambda^4 y \left( v^2 (v^2 -vm +u) +3 \Lambda^4 \right) + \Lambda^{12}=0.}
The antisymmetric representation of $SU(2)$ is just the singlet.
Since a singlet will decouple from the theory, the curve \AStwo\
should reproduce the physics of pure $SU(2)$ Yang-Mills.
Let us set first $m=0$. The condition $p(v)=q(v)$ is then
satisfied and from the way we fixed $A$ our curve reduces to that of
pure $Sp(2)=SU(2)$ gauge theory. We consider now the limit in which
$m$ is sent to infinite. After shifting $v \rightarrow v-m/2$ and
rescaling $y \rightarrow y m^2$, in this limit our curve becomes
\eqn\ASMtwo{y \, (y^2 + y \, (v^2 +u) + \Lambda^4)=0.}
The second factor is precisely the Seiberg-Witten curve for
pure $SU(2)$ gauge theory \swi\ \swii\ \klty\ \arfa.

For $0 < m <  \infty$ \AStwo\ does not factorizes. However the fact
that we can take $m \rightarrow \infty$ without having to tune
the value of the dynamical scale shows that the field of mass $m$ does
not contribute to the beta-function, and therefore
can not be charged under the gauge group.
Furthermore we can calculate the discriminant\foot{We mean here the
loci in moduli space at which \AStwo\ acquire a non-generic singularity,
i.e. excluding the singularity always occurring at $v=0$.} of \AStwo
\eqn\distwo{\Delta=m(u-{m^2 \over 4} +4{\tilde \Lambda}^2)
(u-{m^2 \over 4} -4{\tilde \Lambda}^2 )(u^3-27 m^2 {\tilde \Lambda}^4).}
The first factor is associated with the factorization \antifact.
We will assume that the cycle shrinking at this singularity is a
non physical one, as was the case for the curve \softly\ of the previous
section. If we shift $u \rightarrow u + {m^2 \over 4}$ the second and
third factors match with the discriminant of $SU(2)$ without matter.
We leave for the end of this section the analysis of the last factor.
We will argue that it governs a non-physical singularity
and therefore we can discard it.

We study next the case $N=3$. For $SU(3)$ the antisymmetric
representation coincides with the anti-fundamental. $\CN=2$
hypermultiplets will contain simultaneously fields
transforming in the anti-fundamental and fundamental representation.
Thus \AS\ for $N=3$ should reproduce $\CN=2$ $SU(3)$ gauge
theory with one hypermultiplet in the fundamental.
The known curve for this case is \haoz\ \aps\
\eqn\suthree{y^2 + y \, (v^3 + Wv + U) + 4(v+m)=0,}
where we have set arbitrarily the dynamical scale $\Lambda'$ to
${\Lambda'}^5=4$. The discriminant of \suthree\ is the following rather
lenghthy expression
\eqn\horror{\eqalign{&\Delta=
216000 m^2 U+22680 m^2 U^2 W+13680 W^3 m U-729 U^4 W m-24300 U^3 m \cr
& - 200000-10800 W^2 U^2+90000 W U-66000 m W^2-3552 m^2 W^4-259200 m^3 W \cr
& +729 U^5+216 U^3 W^3+16 W^6 U-16 m W^7-16 m^3 W^6-3456 m^4 W^3-186624 m^5
\cr
& -64 W^5-729 U^4 m^3-216 W^4 U^2 m-216 W^3 U^2 m^3+10368 m^3 W^2 U
+23328 m^4 U^2.\cr}}
Using Maple we could calculate the discriminant of the proposed curve
\AS\ with $p(v)=v^3 + {3m \over 2} v^2 +wv +u$, for the following
cases: {\it i)} $(m=0,u,w)$; {\it ii)} $(m,u=0,w)$; {\it iii)} $(m,u,w=0)$.
After fixing $\Lambda=1$ and relating $(U,W)$ and $(u,v)$ by
\eqn\hev{\eqalign{&  U=u -{1 \over 2} w m +  {1 \over 4} m^3, \cr
& W=w-{3 \over 4} m^2 .\cr}}
the obtained discriminant exactly matches \horror\ up to an
additional prefactor
\eqn\prefact{u^3-27 w^2.}
Notice that the identifications \hev\ coincide with what is expected
from the relation between Higgs vev's and positions of fourbranes
explained in section 5.

For arbitrary $N$ let us denote by $u$ and $w$ the highest and next
to highest order Casimirs, $p(v)=v^N+ \dots + w v +u$. The expression
\prefact\ is part of the discriminant for all the curves \AS. This can
be checked more easily after rewriting the curves in the form (7.3),
which is smooth at the origin. In the rest of this section we will
be referring to the curves in that formulation. One can then see that
when \prefact\ equals zero a
singularity forms at $v=0$. In the case of $SU(2)$ the expression we are
analyzing corresponds to the last factor in \distwo.

When $N$ is even we can study the singularity in the following way.
We consider again the situation $p(v)=q(v)$ in which our cubic curves
factorize. One of the pieces reproduces the Seiberg-Witten curves for
symplectic groups. The expression \prefact\ reduces now to
\eqn\presymp{u=0.}
It was shown in \lll\ that this describes a non-physical singularity
of the symplectic curves.
At $u=0$ a cycle of the opposite behavior under the $\IZ_2$ involution
that the Seiberg-Witten differential shrinks to zero size. Since
at the zeroes of \prefact\ only two branch points collide\foot{The branch
points in the $v$-plane are located where \AS\ has double points as
a polynomial in $y$.}, the associated
vanishing
cycle must be non-physical also when we move to a generic situation
$p(v) \neq q(v)$. Therefore we can discard this singularity for $N$
even.

Let us redefine $u=3 \rho^2 + \epsilon$ and $w=\rho^3$. Since the
singularity governed by \prefact\ develops at $v=0$ for all $N$,
in order to analyze it we could approximate $p(v) \sim w v +u$
for sufficiently small $\epsilon$. In an small neighborhood around
$v=0$, for sufficiently small $\epsilon$, all the curves behave
then as the case $N=2$. For $N$ even we could show that \prefact\
is associated with a non-physical singularity. Based in the local
form of the curve close to the degenerate situation, we can
extend this conclusion to all $N$.

\newsec{Relation to the Atiyah-Hitchin Space}

We have used the semi-classical description of the orientifold
\eqn\AHS{xy=v^{-4}}
as a tool that allowed us to propose new Seiberg-Witten
curves for four-dimensional $\CN=2$ gauge theories. Although this
description is not exact, the key point is that it allowed for additional
factors in the curves proportional to negative powers of $v$. Using
these factors we were able to take into account strong coupling
effects that our brane probes are feeling but are not included
in \AHS. In this section we will study the relation between the
obtained curves and the exact description of the orientifold, the
Atiyah-Hitchin space.

After fixing the coefficients $A$ and $B$ in \antisym, we redefined
$y\rightarrow y v^{-2}$ in order to eliminate the negative powers
of $v$. This led to the curves \AS, which have a
singularity at the origin for any value of the parameters.
Let us consider instead the redefinition $y\rightarrow y v^{-1}$,
we then have
\eqn\Antisym{y^3 + y^2 \, \left( v p(v) + 3 v^{-1} \right)  +
y \, \left( q(v) + 3 v^{-2} \right)  + v^{-3} =0.}
We can eliminate all negative powers in $v$ by shifting $y \rightarrow
y - v^{-1}$, obtaining
\eqn\ASR{y^3 + v \, y^2 p + y \, (q - 2p )  +
v^{-1}(p-q) =0.}
Notice that $q(v)=p(-v)$ and then $p-q=v f(v^2)$. This form of
the curve is regular at $v=0$.

Substituting the previous redefinition of $y$ and an analogous
one for $x$ in \AHS\ we get the space $xy - (x+y) v^{-1}=0$.
It is convenient to change once more coordinates to $({\tilde x},
{\tilde y})$ defined by $y={\tilde x}$, $ x=-{\tilde x}- v {\tilde y}$.
We then obtain the space
\eqn\QAH{{\tilde x}^2+ v {\tilde x} {\tilde y} - {\tilde y}=0.}
which contains no negative powers of $v$ and is smooth at $v=0$.
Equation \ASR\ can be seen as a curve in this space after
just rewriting ${\tilde x}$ instead of $y$. We can use \QAH\ to
eliminate the highest order in $y$ from these curves. This
greatly simplifies \ASR, reducing it to the equation
\eqn\niceA{{\tilde x}(p -{\tilde y}) = v^{-1}(p-q).}

Finally let us shift ${\tilde x} = {\tilde x}' - {v {\tilde y}
\over 2}$ and define $z=v^2$, transforming \QAH\ into
\eqn\AH{{{\tilde x}'}{^2}={ z {\tilde y}^2 \over 4} + {\tilde y}.}
This equation defines the Atiyah-Hitchin space as a complex
manifold in one of its complex structures \threesw\ \athi.
The coordinates $(z,{\tilde x}', {\tilde y})$ are invariant
under the $\IZ_2$ transformation imposed by the orientifold,
that in the initial variables read $(v,y,x) \rightarrow
(-v,x,y)$.
We substitute now ${\tilde x}'$ into \niceA\ and use
\AH\ in order to eliminate this variable. The result is
\eqn\ASAH{{\tilde y}({\tilde y}-p)({\tilde y}-q)=v^{-2}(p-q)^2.}
It is immediate to see that this equation only contains even
powers of $v$ and therefore defines a Riemann surface
$F(z,{\tilde y})=0$ in the complex manifold \AH.
What we have achieved by going through all the coordinate changes
is the following. We have translated the strong coupling effects
encoded in the constants $A$ and $B$ of \antisym\ into the
background geometry describing the orientifold. In this way we
have replaced the approximate description of the orientifold
\AHS\ by the exact one, valid for all $v$. The key steps were given
in \Antisym\ and \ASR. There it was crucial having $A$ and $B$ set
to the values deduced in the past section mostly from field theory
arguments. This shows once more the remarkable interplay between
gauge theory, and strings and M-theory physics.

Before ending this section, let us notice that when the odd powers
of $v$ are switched off in the polynomial $p$ we have $p=q$ and
equation \ASAH\ factorizes. The first factor
${\tilde y}=0$ maps to the first factor in \antifact. For the
configuration of three fivebranes and fourbranes
we are describing, this corresponds to the decoupling of the
middle fivebrane. The rests reduces to
\eqn\SAH{{\tilde y}-p(z)=0.}
Equation \SAH\ was presented in \vafathree, derived from local
mirror symmetry \kkv, for describing the Seiberg-Witten curves
of symplectic gauge groups immersed in $D_n$ spaces. $D_0$
denotes the complex manifold \AH. By just substituting \SAH\
into it we recover the curves for pure Yang-Mills as
formulated in \arsh. We could have obtained \SAH\ by directly
beginning with the curve \symp\ and follow exactly the same
steps presented in this section.

Equation \AHS\ is associated with an orientifold sixplane of charge
$-4$ in the absence of sixbranes. We can include sixbranes by
considering the space $xy=v^{-4} \prod_{a=1}^d (v^2 - m_{a}^2)$.
A direct extension of the considerations in the past section will
lead then to the curves for $SU(N)$ gauge theories with $d$
hypermultiplets in the fundamental representation and one in the
antisymmetric. Following an analogous chain of coordinate changes done
in this section, we would have obtained the defining equation of a
$D_n$ space instead of the $D_0$ space \AH.

\newsec{Product Groups: $SU(N) \times Sp(2k)$}

We will consider briefly a brane configuration containing
four fivebranes in the presence of the negative charge orientifold
sixplane. We will place $N$ fourbranes between the first and second
fivebrane and $2k$ between the second and third. The rest follows
from the $\IZ_2$ invariance imposed by the orientifold (see \product).
This configuration induces an $SU(N)\times Sp(2k)$ gauge theory
on the four macroscopic dimensions of the fourbranes. The matter
content consists in a $\CN=2$ hypermultiplet transforming of the
bifundamental $({\bf N}, {\bf 2k})$.

Again we will use the space \AHS\ as a tool for deriving the curve
describing this configuration. Applying \sup\ to our new case, we
obtain
\eqn\susp{\eqalign{& y^4 + y^3 \, \left( p(v) + D v^{-1} + C v^{-2}
\right)  \cr
& + y^2 \, \left( b(v^2) + B v^{-2} + A v^{-4} \right)
+ v^{-4} y \, \left( q(v) -  D v^{-1} + C v^{-2} \right)  + v^{-8}=0,  \cr}}
where $p$ is a generic polynomial of order $N$, $q(v)=p(-v)$
and $b$ is a polynomial in $v^2$ of degree $k$. The polynomials
$p$ and $b$ include a total of $N+k+2$ parameters. $N+k-1$ of
them will represent expectation values for the $Sp(2k)$ and $SU(N)$
Casimirs, and one will correspond to a bare mass term for the
$({\bf N}, {\bf 2k})$ hypermultiplet.
Another one is related to the ratio between the coupling constants
of both factor gauge groups and the last one can be eliminated
by rescaling $v$. Thus we have the right number of parameters
for describing the Coulomb branch of the product gauge theory.

We could try to fix the constants $A,B,C,D$ by arguments based on
gauge theory. However we will just use the criteria that
\susp\ can alternatively be written as a curve in the
Atiyah-Hitchin space. This uniquely fixes all the constants.
In order to make the structure of the solution more explicit we
reintroduce a constant ${\tilde \Lambda}$ such that
$xy={\tilde \Lambda}^2 v^{-4}$. We write the resulting curve after
rescaling $y \rightarrow v^{-2} y$
\eqn\susp{\eqalign{& y^4 + y^3 \left( e v^2 \prod_{i=1}^N (v-a_i)+ 4
{\tilde \Lambda}
\right)   + y^2 \left( v^4 \prod_{j=1}^k (v^2-b^{2}_i) + 2 e
{\tilde \Lambda} v^2
\prod_{i=1}^N (-a_i)+ 6 {\tilde \Lambda}^2 \right)  \cr
& \;\;\;\;\;\;\;\;\;\;\;\;\; + {\tilde \Lambda}^2 y \left( (-)^N e v^2
\prod_{i=1}^N (v+a_i)+ 4 {\tilde \Lambda} \right)  +
{\tilde \Lambda}^4 =0.  \cr}}

As a consistency check of this curve we analyze the limit in
which the first and the fourth fivebranes are moved off to infinity.
This is achieved by rescaling $y \rightarrow e^{-1} y$ and
sending $e \rightarrow \infty$ and ${\tilde \Lambda} \rightarrow 0$ in such
a way that their product remains finite. Notice that the parameter
$e$ represents the ratio between the coupling constants of the
$SU(N)$ and $Sp(2k)$ groups. The coupling constant of the $SU(N)$
factor is inversely proportional to the distance between
the first and second fivebranes. The limit $e \rightarrow \infty$
corresponds to send it to zero, decoupling the gauge degrees of freedom
of this group factor. In this limit the previous curve reduces to
\eqn\sm{ y^2 \prod_{i=1}^N (v-a_i) + y \left( v^2 \prod_{j=1}^k (v^2-b^{2}_i)
+ 2 e {\tilde \Lambda} \prod_{i=1}^N (-a_i) \right) + (-)^N
(e {\tilde \Lambda})^2 \prod_{i=1}^N (v+a_i) =0,}
This curve is equivalent to the known curve describing an $\CN=2$ gauge
theory with group $Sp(2k)$ and $N$ hypermultiplets in the fundamental
representation \arsh.

\vskip2cm

{\bf Acknowledgments}

The research of K. L. is supported by the Fonds zur F\"orderung
der wissenschaftlichen Forschung, Erwin Schr\"odinger
Auslandsstipendium J01157-PHY and by DOE grant DOE-91ER40618.
E.L. is supported by a C.A.P.V. fellowship.

\listrefs
\end